\documentclass[epj]{svjour}
\usepackage{graphics}
\begin{document}
\title{Electromagnetic Transition Form Factors of Nucleon Resonances. }
\author{Volker D. Burkert } 
\institute{Jefferson Laboratory \\12000 Jefferson Avenue, Newport News, VA23606 \\
E-mail: burkert@jlab.org}
\date{\today}
\twocolumn
\abstract{
Recent measurements of nucleon resonance transition form factors with CLAS at Jefferson
 Lab are discussed.
The new data resolve a long-standing puzzle of the nature of the  
Roper resonance, and confirm the assertion of the symmetric constituent quark 
model of the Roper as the first radial
excitation of the nucleon. The data on high $Q^2$ $n\pi^+$ production 
confirm the slow fall off of the $S_{11}(1535)$ transition form factor with 
$Q^2$, and better constrain the branching ratios $\beta_{N\pi}=0.50$ and 
$\beta_{N\eta}=0.45$. For the first time, 
the longitudinal transition amplitude to the $S_{11}(1535)$ was extracted 
from the $n\pi^+$ data. Also, new results on the transition 
amplitudes  for the $D_{13}(1520)$ resonance are presented showing a 
rapid transition from helicity 3/2 dominance seen at the real photon point 
to helicty 1/2 dominance at higher $Q^2$.}

%
\maketitle

\section{Introduction}
\label{intro}

Electroexcitation of nucleon resonances has long been recognized as a sensitive tool
in the exploration of the complex nucleon structure at varying distances scales.  
Resonances play an important role in fully understanding the spin structure of the nucleon. 
More than 80\%  of the helicity-dependent integrated total photoabsorption cross section 
difference (GDH integral) are the result of the $N\Delta(1232)$ transition \cite{buli,mami}, 
and at a photon virtuality $Q^2=1$~GeV$^2$ more than 50\% of the first moment 
$\Gamma_1^P(Q^2) = \int_0^1{g_1(x,Q^2)dx}$ of the spin structure function $g_1$ for the 
proton are due to contributions of the resonance region at $W<2$~GeV \cite{fatemi03}, 
and are crucial for describing the entire $Q^2$ range of $\Gamma_1^p(Q^2)$
and $\Gamma_1^{p-n}(Q^2)$ for the proton and proton-neutron difference 
respectively \cite{prok08,ioffe,deur}.

\begin{figure}[top]
\resizebox{0.50\textwidth}{!}
{\includegraphics{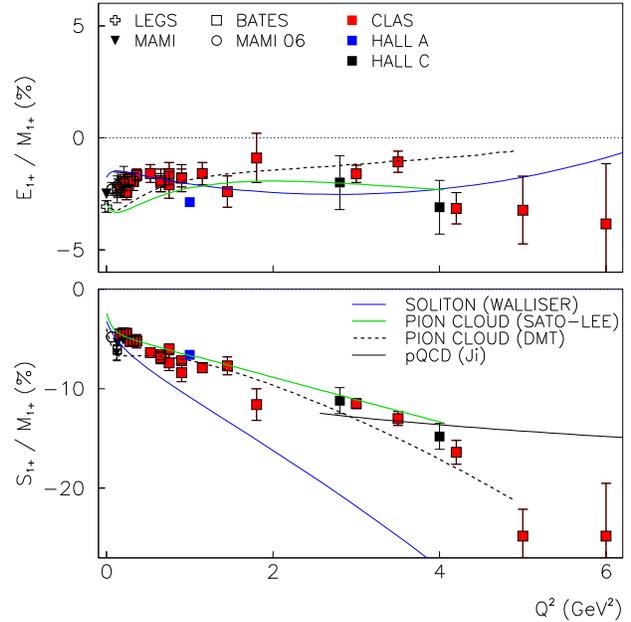}}
\caption{\small $R_{EM}$ and $R_{SM}$ extracted from exclusive reactions $p(e,e^\prime p)\pi^0$ 
using modern analysis tools, e.g. unitary isobar models and dispersion relations.  
Recent  quenched Lattice QCD points are shown as well.}
\label{remrsm}
\end{figure}

Nucleon resonances are of high interest in their own rights. Electroexcitation 
of resonances allows us to probe the internal structure of the excited state 
knowing the structure of the ground state. The most comprehensive predictions 
of the resonance excitation spectrum come from the various implementation 
of the symmetric constituent quark model based on broken $SU(6)$ 
symmetry \cite{isgur}. Other models predict a different excitation spectrum, 
e.g. through a diquark-quark picture, or through 
dynamical baryon-meson interactions. The different resonance models not only 
predict different excitation spectra but also different $Q^2$ dependence of 
transition form factors.  Mapping out the transition form factors will tell 
us a great deal about the underlying quark or hadronic structure.    

CLAS is the first full acceptance instrument with sufficient resolution to measure exclusive 
electroproduction of mesons with the goal of studying the excitation of nucleon resonances in 
detail. The entire resonance mass region, a large range in the photon virtuality $Q^2$ can be 
studied, and many meson final states are measured simultaneously \cite{burkert-lee}. 
In this talk I discuss recent 
results from the electroproduction of single pions to study several well-known excited states. 

\section{The $N\Delta(1232)$ transition}
\label{sec:ndelta}
An interesting aspect of nucleon structure at low energies 
is a possible quadrupole deformation of the nucleon's lowest excited state, the 
$\Delta(1232)$. 
Such a deformation would be evident in non-zero values of the quadrupole transition 
amplitude $E_{1+}$ from the nucleon to the $\Delta(1232)$ \cite{buchmann}. 
In models with $SU(6)$ spherical symmetry, the $N\Delta$ transition is simply due to
a magnetic dipole $M_{1+}$ mediated by a spin flip, and $E_{1+} = S_{1+} = 0$ . 
Dynamically, quadrupole deformations may arise through the interaction of the photon with the pion 
cloud \cite{sato,yang} or through the one-gluon exchange mechanism \cite{koniuk}. 
At asymptotic momentum transfer, a model-independent prediction of helicity conservation 
requires $R_{EM}\equiv E_{1+}/M_{1+} \rightarrow +1$. An interpretation of $R_{EM}$ in 
terms of a quadrupole deformation can thus only be valid at low momentum transfer.  
    
\begin{figure}
\resizebox{0.50\textwidth}{!}
{\includegraphics{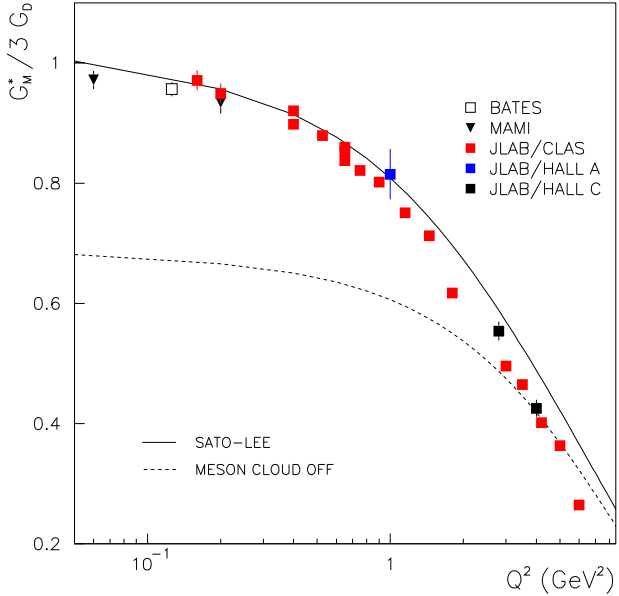}}
\caption{\small Magnetic transition form factor $GM^\Delta$ extracted from exclusive
reactions $p(e,e^\prime p)\pi^0$, and normalized to the dipole form.}
\label{gm}
\end{figure}

Results of the multipole analysis of the JLab data \cite{kjoo,frolov,ungaro,kelly} as well 
as low $Q^2$ data from MAMI \cite{mami-delta,beck}, Bates \cite{bates-delta} and LEGS \cite{legs}
are shown in Fig.\ref{remrsm}.  A consistent picture emerges from these precise data. 
\begin{itemize}
\item{} $R_{EM}$ remains negative, small and nearly constant in the entire range $0< Q^2<6$~GeV$^2$.
\item{} There are no indications that leading pQCD contributions are important as 
they would result in $R_{EM} \rightarrow +1$ \cite{carlson}. 
\item{} $R_{SM}$ also remains negative, but its magnitude strongly rises with $Q^2$.
\end{itemize} 
Comparison with microscopic models shows that simultaneous description of 
both $R_{EM}$ and $R_{SM}$ is achieved with dynamical models that include pion-nucleon
interactions explicitly. This supports the claim that most of the 
quadrupole strength in the $N\Delta(1232)$ transition is due to meson effects which 
are usually not included in quark models. From Fig.~\ref{gm} we conclude that at the 
real photon point 1/3 of the transition strength is due to pion effects, which extends
to rather high $Q^2$, although with decreasing relative strength.  

The MAID unitary isobar model has been frequently used in the analysis of pion electroproduction data. 
I want to comment on one aspect of the 2007 version MAID07 that has generated some confusion 
regarding the results of  analysis compared to the 2003 version MAID03. 
\begin{figure}[top]
\resizebox{0.5\textwidth}{!}
{\includegraphics{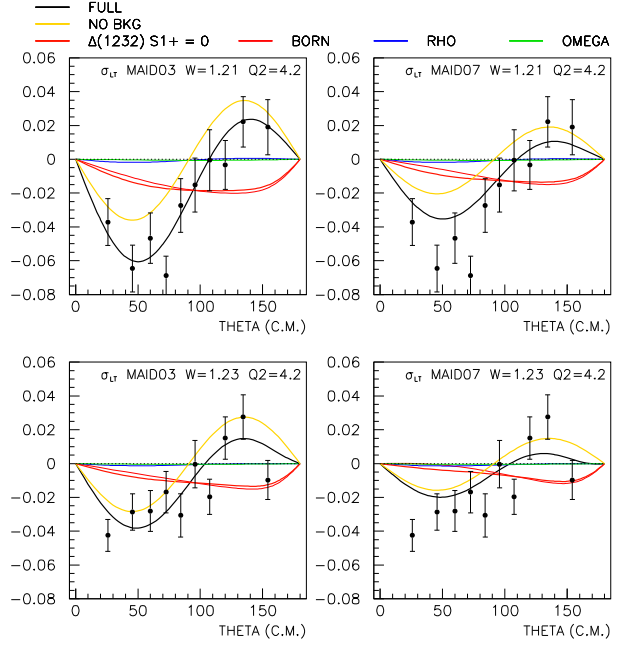}}
\caption{\small Angular dependence of response function $\sigma_{LT}$ from 
the high $Q^2$ CLAS data. 
extracted from exclusive reactions $p(e,e^\prime p)\pi^0$ in the $\Delta(1232)$ 
region. The curves represents MAID03 and MAID07 parameterizations. }
\label{m07m03} 
\end{figure}
Independent analyses of the JLab data from CLAS and Hall C have been carried out within 
the MAID framework of MAID03 \cite{maid03} and MAID07 \cite{maid07}. MAID03 parameters 
were adjusted by analysing the 2001 CLAS cross section data \cite{kjoo} and the 
higher $Q^2$ Hall C data \cite{frolov}. In Fig.~\ref{m07m03} the MAID03 parameters are 
used and the predictions compared with the CLAS data published in 2006 \cite{ungaro} for the 
response function $\sigma_{LT}$ in the $\Delta(1232)$ mass region showing excellent agreement.
The extracted $R_{SM}$ ratio showed a strong rise in magnitude with $Q^2$  consistent with 
the values shown in Fig.~\ref{remrsm} from the CLAS 2006 data that were not included in the fit.
In MAID07 a new fit was made that now included the CLAS 2006 data but did not include the 
previously used Hall C data. The results of that fit are shown in the right panels in 
Fig.~\ref{m07m03}, 
and clearly compares much less favorably with the measured $\sigma_{LT}$ resonse function.
It also results in an almost $Q^2$-independent behavior at high $Q^2$, in clear contradiction 
to the previously obtained strong rise in magnitude with $Q^2$.  It appears that 
this discrepancy is an artifact of the parameterization used in MAID07 for the $R_{SM}$ ratio, 
which includes the 
constraint $R_{SM}\rightarrow {\rm constant}$, the asymptotic limit for
$Q^2 \rightarrow \infty$. However, this constraint is not justified as there are no 
indications that asymptotic 
behavior is relevant either in $R_{EM}$ (which would require $R_{EM} \rightarrow +1$, 
while the data show $R_{EM} \approx -0.03$), or in the extraction of $R_{SM}$ when not 
constraint by the presumed asymptotic behavior. 

Ultimately, we want to come to a QCD description of these important 
nucleon structure quantities. In recent years significant effort has been  
extended towards a Lattice QCD description of the $N\Delta$ 
transition \cite{alexandrou1,alexandrou2}. Within the still large error bars, both quenched 
and unquenched calculations at
$Q^2<1.5$~GeV$^2$ with pion masses of 400~MeV are consistent with a constant 
negative value of $R_{EM}\approx -0.02$, in agreement with the data.  For the $R_{SM}$ 
ratio there is a clear discrepancy at low $Q^2$ in both quenched and unquenched QCD 
calculation while the rise in magnitude of $R_{SM}$ with $Q^2$ observed in the data 
is quantitatively reproduced in full QCD at the $Q^2>1$GeV$^2$.    

The measured $N\Delta$ transition form factors extend to $Q^2=6$~GeV$^2$, and show 
no sign of the expected asymptotic behavior. It would 
be very interesting to see if LQCD calculations can describe the observed strong 
$Q^2$ dependence of $R_{SM}$, and the near lack of $Q^2$ dependence of $R_{EM}$ at 
high $Q^2$.

\section{The second resonance region}
\label{sec:2ndres}
Three states, the ``Roper'' $P_{11}(1440)$, 
and two strong negative parity states, $D_{13}(1520)$, 
and $S_{11}(1535)$ make up the 
second enhancement seen in inclusive electron scattering.

\begin{figure}[here]
\resizebox{0.48\textwidth}{!}{\includegraphics{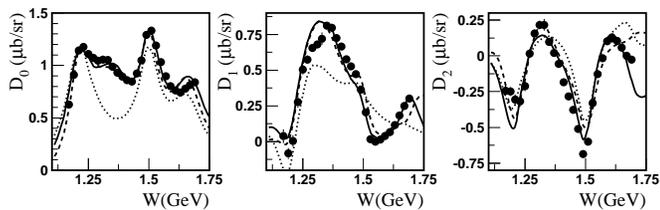}} 
\caption{\small\protect W dependence of the three lowest Legendre moments from $n\pi^+$ 
angular distributions at fixed $Q^2=2.05$~GeV$^2$. The dotted line indicates the cross section
when the amplitudes of the $P_{11}(1440$ are set equal 0.}
\label{legendre}
\end{figure}

\subsection{The Roper resonance $P_{11}(1440)$ - a puzzle resolved }

The $P_{11}(1440)$ resonance has been a focus of attention for the last decade, largely due
to the inability of the standard constituent quark model to describe basic
features such as the mass, photocouplings, and $Q^2$ evolution. 
This has led to alternate approaches where the state is treated as 
a gluonic excitation of the nucleon \cite{libuli}, or has a small quark core 
with a large meson cloud \cite{cano}, or is a hadronic molecule of a nucleon and a 
$\sigma$ meson \cite{krewald}.
Quenched lattice QCD calculations \cite{lattice-1} indicate that the 
state has a significant 3-quark component, and calculate the mass to be 
close to the experimental value.

Given these different theoretical concept for the structure of the state, 
the question ``what is the nature of the Roper state?'' has been a focus of 
the $N^*$ program with CLAS.  The state couples to both $N\pi$ and $N\pi\pi$ 
final states. It is also a very wide resonance with about 350 MeV total width. 
Therefore single and double pion electroproduction data covering a large range in the 
invariant mass W, with full center-of-mass angular coverage are crucial in extracting the 
transition form factors in a large range of $Q^2$. As an isospin  $I = {1\over 2}$ state, 
the $P_{11}(1440)$ couples more strongly to n$\pi^+$ than to p$\pi^o$. Also
contributions of the high energy tail of the $\Delta(1232)$ are much reduced in that 
channel due to the $I = {3\over 2}$ of the $\Delta(1232)$. 

\begin{figure}[top]
\resizebox{0.48\textwidth}{!}{\includegraphics{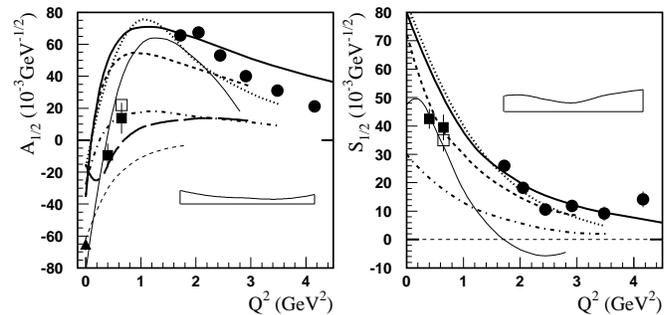}} 
\caption {\small Transverse electrocoupling amplitude for the 
Roper $P_{11}(1440)$ (left panel). The full circles 
are the new CLAS results. The squares are previously published results of fits to CLAS data
at low $Q^2$. The right panel shows the longitudinal amplitude.} 
\label{roper}
\end{figure}

Over 33,000 differential cross section data points and polarized beam asymmetries from 
CLAS \cite{park08} have
been analyzed using a fixed-t dispersion relations approach and a unitary isobar model (UIM)~\cite{janr}. Some of the features of the data may best be seen in the Legendre moments. 
Response functions can be expressed in terms of Legendre polynomials, e.g.
the azimuthal angle independent part of the differential cross section can be written as:
$$\sigma_T + \epsilon \sigma_L = \sum_{\ell=0}^\infty D_\ell^{T+L}P_\ell({cos\Theta^*_\pi}).$$  
Figure~\ref{legendre} shows the lowest Legendre moments for this response functions. 
The transverse and longitudial electro-coupling amplitudes $A_{1/2}$ and $S_{1/2}$ of 
the transition to the $P_{11}(1440)$ resonance are extracted from fits to the 
data \cite{aznauryan-2}. They are shown in Fig.~\ref{roper}.      
At the real photon point $A_{1/2}$ is negative. The CLAS results show a fast rise of the 
amplitude with $Q^2$ and a sign change near $Q^2=0.5$~GeV$^2$. At $Q^2=2$GeV$^2$ the 
amplitude has about the same magnitude but opposite
sign as at $Q^2=0$. It slowly falls off at high $Q^2$. This remarkable behavior of a sign 
change with $Q^2$ has not been seen before for any nucleon transition form factor or 
elastic form factor. The longitudinal amplitude $S_{1/2}$ is large at low $Q^2$ and drops 
off smoothly with increasing $Q^2$.  The bold curves are all relativistic light front 
quark model calculations \cite{aznauryan-qm}. The thin solid 
line is a non-relativistic quark model with a vector meson cloud \cite{cano}, and 
the thin dashed line is for a gluonic excitation \cite{libuli}. 
The first results for the transition form factors of the Roper have recently been obtained 
in unquenched QCD\cite{roper-lqcd}.

The hybrid baryon model is clearly ruled out for both amplitudes. At high $Q^2$ both 
amplitudes are qualitatively described by the 
light front quark models, which strongly suggests that the Roper is indeed a radial excitation 
of the nucleon. The low $Q^2$ behavior is not well described by the LF models and they
fall short of describing the amplitude at the photon point. This indicates that important
contributions, e.g. meson-baryon interactions at large distances may be missing. 

\subsection{The $S_{11}(1535)$ state}

The $S_{11}(1535)$ state was found to have an unusually 
hard transition formfactor, i.e. the $Q^2$ evolution shows a slow
fall-off. This state has mostly been studied in the $p\eta$ channel where the
$S_{11}(1535)$ appears as a rather isolated resonance near the $N\eta$ threshold 
and with very little non-resonant background.
Data from JLab using CLAS \cite{thompson,denizli} and Hall C \cite{armstrong} 
instrumentation, have provided a consistent picture of the $Q^2$ evolution obtained 
from $\eta$ electroproduction data alone, confirming the hard form factor 
behavior with  precision. 
There are two remaining significant uncertainty in the electromagnetic couplings 
of the $S_{11}(1535)$ that need to be examined. The first one is due to the 
branching ratio of the $S_{11}(1535) \rightarrow p\eta$, the second one is due 
to the lack of precise information on the longitudinal coupling, which in the $p\eta$ channel
is usually neglected. 

\begin{figure}
\resizebox{0.45\textwidth}{!}{\includegraphics{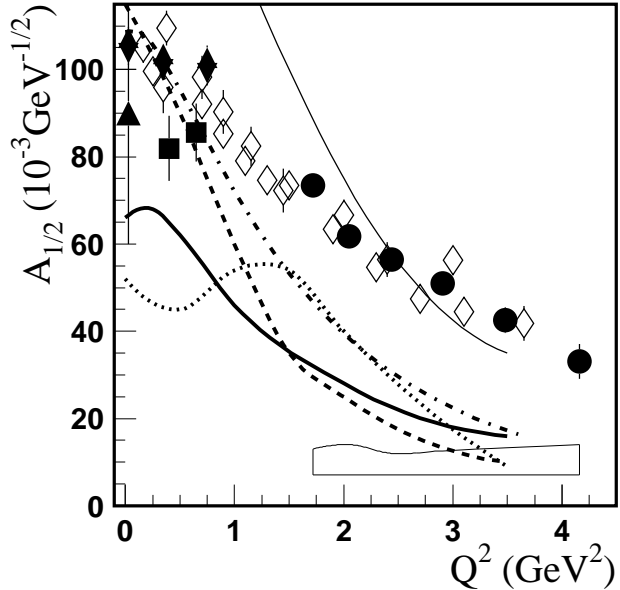}} 
\caption {\small The transition amplitude $A_{1/2}$ for the $S_{11}(1535)$. See text for 
explanations.}
\label{s11}
\end{figure}

The $p\eta$ data have been normalized 
using a branching ratio $\beta_{N\eta} = 0.52$, while the PDG gives a range of 
$\beta^{PDG}_{N\eta} = 0.45 - 0.60$. 
Since this state practically does not couple to other channels than $N\eta$ and $N\pi$, 
a measurement of the reaction $ep\rightarrow e\pi^+n$ will reduce this uncertainty.
Also, the $N\pi$ final state is much more sensitive to the longitudinal amplitude due to
a strong $S_{11}-P_{11}$ interference term present in the $N\pi$ channel. 
With these goals in mind the CLAS $n\pi^+$ data have been used to 
determine the electrocoupling amplitudes for the $S_{11}(1535)$. 
Using the average values $\bar{\beta}^{PDG}_{N\pi}=0.45$ and $\bar{\beta}^{PDG}_{N\eta}=0.52$, 
the $n\pi^+$ data fall systematically above the $p\eta$ data set. 
Adjusting $\beta_{N\pi} = 0.50$ and $\beta_{N\eta} = 0.45$ brings the two data sets into 
excellent agreement for the higher $Q^2$ data, as shown in Fig.~\ref{s11}. The diamond 
symbols show the results in the $p\eta$ channel. The full circles are from the analysis 
of the CLAS $n\pi^+$ data \cite{aznauryan-2}. The square symbols are from the 
analysis of earlier CLAS $p\pi^0$, and $n\pi^+$ data \cite{aznauryan-1} with adjusted 
$\beta_{N\pi}$. The theory curves are from various constituent quark models quark model \cite{capstick,salme,merten,santopinto,warns}. 
There could be a 10-20\% difference between the $N\pi$ and $N\eta$ for the 
$Q^2 = 0.4,~0.6$~GeV$^2$ points. This indicates that meson-cloud effects may play some 
role at low $Q^2$, possibly affecting the results differently in the two channels. Analyses 
that take coupled channel effects into account are needed to fully clarify the 
low $Q^2$ behavior. 
As mentioned above an advantage of the $p\pi$ channel in studying the $S_{11}(1535)$ is 
that it is also 
sensitive to the longitudinal transition amplitude, while the $N\eta$ channel has little 
sensitivity and requires a Rosenbluth separation to separate the transverse and longitudinal 
terms.  In the $N\pi$ case, the sensitivity is due to a significant $s-p$ wave interference 
with the nearby $p$-wave amplitude of the $P_{11}(1440)$. This can be seen in the multipole 
expansion of the lowest Legendre moment for the $\sigma_{LT}$ response function:
 $$D_0^{LT} = {{|\vec{q}|} \over K} Re(E_{0+}S_{1-}^* + S_{0+}M_{1-}^*).$$ The second term is very 
sensitive to the $S_{0+}$ multipole of the $S_{11}(1535)$ due to the strong transverse 
Roper multipole ($M_{1-}$), especially at high $Q^2$. Preliminary results show significant
negative values for the $S_{1/2}$ amplitude of $S_{11}(1535)$.     

\subsection{Helicity structure of the $D_{13}(1520)$ }

\begin{figure}[top]
\resizebox{0.48\textwidth}{!}{\includegraphics{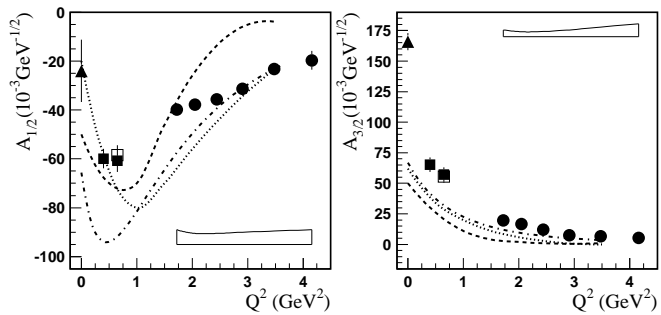}}
\caption{\small Transverse helicity amplitudes $A_{1/2}$ (left panel) and $A_{3/2}$ 
(right panel) for the $D_{13}(1520)$. The helicity switch is clearly visible. 
Model curves as in Fig.6.} 
\label{d13}
\end{figure}

A longstanding prediction of the dynamical constituent quark model is the rapid helicity 
switch from the dominance of the $A_{3/2}$ at the real photo point to the dominance of 
the $A_{1/2}$ amplitude at $Q^2 > 1$~GeV$^2$. In the simple non-relativistic harmonic 
oscillator model with spin and orbit flip amplitudes only, the ratio of the two amplitudes 
is given by: 
$$ {A_{1/2}^{D13} \over A_{3/2}^{D13}} = {-1 \over \sqrt{3}}({\vec{Q}^2 \over \alpha} - 1)~,$$ 
where $\alpha$ is a constant adjusted to reproduce the ratio at the photon point where $A_{1/2}$ is 
very small. Is is clear that the model predicts a rapid rise of the ratio with $Q^2$. Figure~\ref{d13}
shows the results for the two transverse amplitudes. We see the $A_{3/2}$ amplitude decreasing rapidly 
in strength with increasing $Q^2$. The $A_{1/2}$ amplitude increases rapidly in magnitude 
with increasing $Q^2$, before falling off slowly at $Q^2>1$~GeV$^2$. $A_{1/2}$ completely dominates at 
$Q^2>2$~GeV$^2$.

\section{Conclusions} 
\label{sec:conclusions}
With the recent precise data on pion and eta electroproduction, combined with the 
large coverage in $Q^2$,  W, and center-of-mass angle, the study of nucleon resonance 
transitions has become an effective tool in the exploration of nucleon structure 
in the domain of strong QCD and confinement. We have learned that the $\Delta(1232)$ 
exhibits an oblate deformation. The multipole ratios $R_{EM}$ an $R_{SM}$ show no sign
of approaching the predicted asymptotic behavior, which provides a real challenge for model
builders. The latest data from CLAS on charged pion production reveal a sign change 
of the transverse  amplitude for the N-Roper transition near $Q^2 = 0.5$~GeV$^2$, 
and give strong evidence for this state as the first radial excitation of the nucleon. 
The hard transition form factor 
of the $S_{11}(1535)$ previously observed only in the $p\eta$ channel is confirmed in 
the $n\pi^+$ channel, which also allows us to extract the so far unmeasured longitudinal 
amplitude $S_{1/2}$. The $D_{13}(1520)$ clearly exhibits the helicity flip 
behavior long ago predicted by the constituent quark model.

\end{document}